\documentclass[12pt]{article}
\usepackage{a4wide}
\usepackage{amssymb}
\usepackage{graphicx}
\begin{document}
{\renewcommand{\thefootnote}{\fnsymbol{footnote}}
\begin{center}
{\LARGE  (Loop) quantum gravity and the inflationary
  scenario\footnote{Published in {\em Comptes Rendus Physique},
    doi:10.1016/j.crhy.2015.08.007}}\\
\vspace{1.5em}
Martin Bojowald\footnote{e-mail address: {\tt bojowald@gravity.psu.edu}}
\\
\vspace{0.5em}
Institute for Gravitation and the Cosmos,\\
The Pennsylvania State
University,\\
104 Davey Lab, University Park, PA 16802, USA\\
\vspace{1.5em}
\end{center}
}

\setcounter{footnote}{0}

\begin{abstract}
  Quantum gravity, as a fundamental theory of space-time, is expected to
  reveal how the universe may have started, perhaps during or before an
  inflationary epoch. It may then leave a potentially observable (but probably
  minuscule) trace in cosmic large-scale structures that seem to match well
  with predictions of inflation models. A systematic quest to derive such tiny
  effects using one approach, loop quantum gravity, has, however, led to
  unexpected obstacles. Such models remain incomplete, and it is not clear
  whether loop quantum gravity can be consistent as a full theory. But some
  surprising effects appear to be generic and would drastically alter our
  understanding of space-time at large density. These new high-curvature
  phenomena are a consequence of a widening gap between quantum gravity and
  ordinary quantum-field theory on a background.
\end{abstract}

How much of quantum gravity can be tested by cosmological observations?  This
question is non-trivial, not only because the understanding of ``quantum
gravity'' depends on the approach one takes. A second, and perhaps more
important, difficulty consists in the fact that several key aspects of
physical properties usually associated with theories of gravity play
interconnected roles in this context. Based on the lessons of general
relativity, a quantum theory of gravity, in general terms, is expected to
entail (i) a modified dynamics, given by a gravitational force with quantum
corrections; (ii) new quantized modes, gravitons; and (iii) a quantum version
of space-time structure.

While the first two aspects are commonly exploited for proposals of
potentially observable consequences in cosmology, the last one is most crucial
at a conceptual level. Only if this subtle issue is addressed can one be sure
that one is testing quantum gravity, as opposed to a quantum field theory of
tensor modes on some curved background: Classical gravity, described by
general relativity, is characterized not only by a certain form of dynamics
but also, and more fundamentally, by a large class of covariance
symmetries. These symmetries characterize the structure of space-time in
classical gravity as a Riemannian manifold of Lorentzian signature. Quantizing
fields on such a manifold is not the same as quantizing space-time itself: The
full (spatial) metric would not be subject to quantum fluctuations. A full
quantization of the metric, on the other hand, requires great care because, in
the absence of a background, it can be changed by a large class of
transformations which must in some way be respected after quantization, or
else the theory would be coordinate dependent and therefore meaningless.

Quantizing highly symmetric theories is always a delicate process, and the
decades of unsuccessful attempts to find a complete and consistent theory of
quantum gravity bear witness to the fact that this theory is no
exception. Quantum field theory on a curved background does not quantize
space-time and its symmetries, and therefore leaves out an important part of
the question of quantum gravity. A common, and often implicit, assumption is
that quantum field theory on curved space-time may be a good approximation to
quantum gravity in relevant cosmological regimes. However, the process of
gravitons seeding a background of tensor modes is a rather strong
quantum-gravity effect; in fact, it is often taken as the best (or only) way
to ``observe'' or ``establish the quantization of gravity'' \cite{EstQG}. If
this is the case, is it legitimate to treat quantum gravity as some version of
quantum field theory on a curved background, bypassing a quantum theory of
curved space-time itself?

The present contribution does not provide a universal answer to this
question. However, an example will be given in which quantum field theory on a
curved background turns out to be not just a poor approximation to a specific
model of quantum gravity, but to result in a different and incompatible
scenario. The model has been formulated in the framework of loop quantum
gravity, following systematic attempts to derive potential observational
consequences of this theory. Loop quantum gravity remains incomplete, and it
is not clear how generally the effects appear that are responsible for the
mismatch between quantum space-time and quantum field theory on a curved
background, even though at present they seem to be rather generic. The result
is sobering and provides much caution against too-optimistic presentations of
potentially observable quantum-gravity effects, not only in loop quantum
gravity itself but perhaps also more generally. This article therefore
refrains from presenting specific details of potential observations in loop
quantum gravity, as they all appear to be largely ambiguous at present. For
some of the known options, interested readers are referred to reviews such as
\cite{GCPheno,ObsLQC}.

The different aspects of quantum gravity, listed above, are all
interrelated. Graviton scattering implies loop corrections in the perturbative
law of the gravitational force. A fundamental version of this law remains to
be found, owing to the problem that gravity is not renormalizable. But in this
context it is sufficient to consider an effective theory of gravity
\cite{EffectiveGR,BurgessLivRev} in which loop corrections to the
gravitational force can be discussed in a well-defined way --- so long as a
background treatment of space-time is valid. The structure of space-time
enters into the considerations because it is closely linked to the modes and
dynamics of gravity. One obtains the Einstein--Hilbert action from the
covariance symmetries of space-time, leaving much less freedom for
interactions compared with other fields (on a given background). An effective
theory of gravitons on a classical background contains the usual
higher-curvature terms.

When one quantizes gravity, the structure of space-time may change, depending
on the specific approach used. Well-known proposals include non-commutative
geometry or discreteness. If space-time is no longer modeled by a Riemannian
manifold, it is not clear how general covariance can still apply in a
meaningful way, but there must be some form in any consistent approach because
the breaking of covariance symmetries would amount to a gauge anomaly. Only
after the form of space-time has been determined in a given approach can one
proceed and compute dynamical quantum corrections. Effective theory, as the
main and most powerful tool to derive reliable quantum effects, relies on
knowledge of the symmetries of a theory. Symmetries determine which terms an
effective action may contain, the coefficients of which one can then compute
by more-detailed calculations. Non-Riemannian structures can lead to
quantum-gravity effects which a treatment of quantum-field theory on a curved
background of classical form would miss.

A canonical approach is often useful in order to reveal non-trivial aspects of
quantum theories. It is of special importance in quantum gravity because it
does not presuppose what structure space-time may have. Symmetries are
represented by generators, which are quantized in a canonical approach and may
have quantum corrections. If also their algebra turns out to be modified (but
not broken for an anomaly-free theory), a new structure of space-time
emerges. The rest of this article attempts to review these features without
too many technical details, followed by an example in which quantum-field
theory on a curved background differs crucially from a model of (loop) quantum
gravity.

In any quantum theory, dynamics is generated by a Hamiltonian operator
$\hat{H}$, in quantum mechanics following the Schr\"odinger equation
$i\hbar \partial\psi/\partial t=\hat{H}\psi$. In a covariant theory, the time
coordinate can be changed freely (and locally), so that $\hat{H}$ must also be
a generator of symmetries. This dual role of Hamiltonian operators is one of
the reasons why quantum gravity is complicated and incomplete, but its
conceptual consequences can be explored by well-known methods.

General covariance in other words is local Lorentz (or Poincar\'e)
invariance. In contrast to a quantum-mechanics situation (or minisuperspace
quantum cosmology) with a single time parameter $t$ and its symmetry of rigid
time translations $t\mapsto t+a$, a covariant theory allows one to change time
independently at different positions as well as non-linearly: $t\mapsto
t'(t,x)$. Correspondingly, there is not just one generator $\hat{H}$ but a
whole family, or a function $\hat{H}(t,x)$. 

\begin{figure}
\begin{center}
\includegraphics[width=5.5cm]{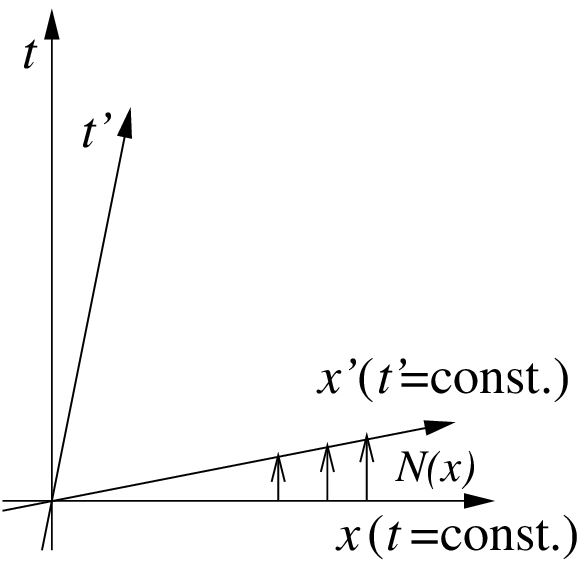}\hspace{1cm}\includegraphics[width=5.5cm]{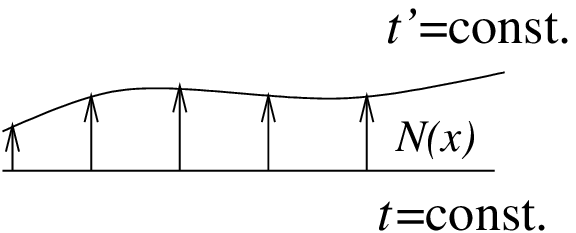}
\caption{An infinitesimal Lorentz transformation shifts points on
  constant-time surfaces in a linear fashion, normal to the surface (left). A
  transformation of general covariance moves points normal to a constant-time
  surface, but non-linearly (right). \label{f:Slices}}
\end{center}
\end{figure}

For transformations $t'(t,x)$ linear in $t$ and $x$, one obtains rigid time
translations and Lorentz boosts, as pictured in Minkowski diagrams familiar
from special relativity; see Fig.~\ref{f:Slices}. The commutator of a time
translation and a boost is a spatial translation, and the commutator of two
boosts is a spatial rotation. The whole Poincar\'e algebra is generated if one
starts with a family $\hat{H}(t,x)$ of Hamiltonians generating linear
space-time transformations.

A generally covariant theory allows one to perform not only local but also
non-linear space-time transformations. The Minkowksi diagrams of special
relativity may then be replaced by arbitrary deformations of curved spatial
slices in space-time (Fig.~\ref{f:Slices}). The position-dependent generator
$\hat{H}(t,x)$ is then conveniently replaced by a ``time deformation''
$T[N]=\int{\rm d}^3x N(x)\hat{H}(t,x)$, where $N$ is a function on the spatial
slice ($t$ constant) so that $T[N]$ signifies an infinitesimal deformation
along the normal direction in space-time, by a distance $N(x)$ at point
$x$. Similarly, spatial deformations within a slice, along vector fields
$\vec{w}$, are denoted as $S[\vec{w}]$. (The function $N$ and vector field
$\vec{w}$ appear as lapse and shift in a canonical formulation of the
gravitational dynamics, such as the ADM decomposition \cite{ADM}. See
\cite{CUP} for details.)

The commutators of such transformations can be derived from the geometry of
spatial 3-manifolds embedded in space-time, resulting in the
hypersurface-deformation algebra \cite{Regained}
\begin{eqnarray}
 [S[\vec{w}_1],S[\vec{w}_2] &=& S[{\cal L}_{\vec{w}_1}\vec{w}_2]\label{SS}\\{}
 [T[N],S[\vec{w}]] &=& -T[{\cal L}_{\vec{w}} N]\\{}
 [T[N_1],T[N_2]] &=& S[ N_1\vec{\nabla}N_2- N_2\vec{\nabla}N_1]\,. \label{TT}
\end{eqnarray}
For linear $\vec{w}$ and $N$ and flat (Euclidean) spatial slices
in Minkowski space-time, one can confirm that the hypersurface-deformation
algebra is reduced to the Poincar\'e algebra.

Unlike the Poincar\'e algebra, however, the hypersurface-deformation algebra
depends on the metric induced by space-time on a spatial slice: The gradients
in (\ref{TT}) are defined only if a metric is provided
($\nabla^aN=h^{ab}\partial N/\partial x^b$ with the inverse spatial metric
$h^{ab}$). Even when space-time is Minkowski but the spatial slices are not
flat do we have a metric entering the algebraic relations, in addition to the
parameters $N$ and $\vec{w}$ of generators. Mathematically, the commutators do
not provide a strict Lie algebra, but a Lie algebroid
\cite{ConsAlgebroid}. Physically, the presence of the inverse metric means
that canonical quantum gravity, which provides operators for the spatial
metric, could lead to quantum corrections in the structure functions of the
hypersurface-deformation algebra. The structure of space-time may then change.

A more-detailed analysis reveals that structure functions of symmetry algebras
are {\em not} subject to quantum corrections in a large number of cases,
provided that the quantum generators can be represented without anomalies
\cite{EffConsQBR}. However, an exception is given by theories that quantize by
not fully removing their regulators. Loop quantum gravity is of this form
because it replaces connection or curvature components $A_a^i$ by
exponentiated versions, so-called holonomies (or parallel transport)
\begin{equation} \label{Hol}
 h_e(A)={\cal P}\exp(\smallint_e \tau_iA_a^i t_e^a{\rm d}s)
\end{equation}
with SU(2)-generators $\tau_i$ and for any spatial curve $e$ with tangent
vector $t^a$ (and path ordering denoted by ${\cal P}$).  Since the local gauge
group is compact, the matrix elements of $h_e(A)$ are bounded functions of
$A_a^i$. Compared with a direct quantization of $A_a^i$ in Wheeler-DeWitt
fashion, holonomies have more convenient mathematical properties as operators
\cite{LoopRep,ALMMT}.

The regulator, given by extended curves used to integrate holonomies or
parallel transport, can be removed without much trouble when one assumes that
the theory is invariant under spatial deformations \cite{AnoFree}. Any
extended curve can then be equivalent to one arbitrarily close to a single
point, and holonomies approximate curvature components exceedingly
well. However, a discussion of anomaly freedom of a quantized
hypersurface-deformation algebra cannot assume invariance under spatial
deformations because these transformations are an important part of the
algebra. The regulator is no longer being removed, and quantum corrections in
the structure functions of the algebra are possible.

At this stage, an effective approach to canonical quantization and symmetries
is useful \cite{EffAc,EffCons}. For decades, it has been an unsolved and
intimidating problem to find quantum generators $\hat{C}_I$ of the
gravitational symmetries in (\ref{SS})--(\ref{TT}) and to compute their
commutators $[\hat{C}_I,\hat{C}_J]$. (As in any theory with local symmetries,
there are infinitely many generators. The index $I$ may therefore lie in a
continuous range.) For a consistent and anomaly-free quantization, the
commutators must form a closed algebra, that is
$[\hat{C}_I,\hat{C}_J]=\hat{f}_{IJ}^K\hat{C}_K$ with suitable structure
operators $\hat{f}_{IJ}^K$ corresponding to
(\ref{SS})--(\ref{TT}). Classically, we have a closed algebra of Poisson
brackets $\{C_I,C_J\}=f_{IJ}^KC_K$ from canonical gravity, but quantization is
sensitive to factor ordering and regularization, so that closure of the
operator algebra is not guaranteed. The canonical effective approach provides
methods by which one can compute a ``quantum Poisson bracket'' defined
as
\begin{equation}
 \{C_I^{\rm   eff},C_J^{\rm eff}\}_{\rm Q}:=
\frac{\langle[\hat{C}_I,\hat{C}_J]\rangle}{i\hbar}\,.
\end{equation}
It is evaluated for effective constraints $C_I^{\rm
  eff}:=\langle\hat{C}_I\rangle$ (computed in generic states) which can be
written as the classical constraints plus an infinite series of quantum
corrections. The crucial observation is that a closed quantum algebra with
structure operators $\hat{f}_{IJ}^K$ implies a closed algebra of effective
constraints to any order of the $\hbar$-expansion of quantum corrections. The
effective Poisson brackets are much easier to compute than operator
commutators. And while the closure of an effective algebra to some order in
$\hbar$ does not imply the existence of a closed operator algebra, a negative
outcome is meaningful: If the effective algebra cannot close to some order in
$\hbar$, it is impossible to find a closed operator version. Or if the
effective algebra can close only if its structure functions have a certain
form of quantum corrections, the operator algebra must be modified as
well. Such algebraic statements are much more general than details of the
precise dynamics computed to some order in $\hbar$: Even if the effective
constraints to some order do not approximate the quantum dynamics well,
potential closed versions of their algebra still have implications for
possible forms of the quantum algebra.

In models of loop quantum gravity, one generic result of this form has been
found \cite{JR,ScalarHolInv,HigherSpatial,DeformedCosmo}: So far, effective
versions of (\ref{TT}) can exist in the presence of holonomy modifications
only if there is an additional function $\beta$ multiplying the generator on
the right-hand side:
\begin{equation}
[T[N_1],T[N_2]] = S[\beta(K)( N_1\vec{\nabla}N_2-
N_2\vec{\nabla}N_1)]\,. \label{TTbeta} 
\end{equation}
The classical structure function (the inverse spatial metric) is therefore
modified. Moreover, if using holonomies modifies the classical curvature
dependence on some component $K$ in the Hamiltonian by replacing the usual
quadratic dependence by a function $f(K)$, then the effective algebra is
closed for $\beta(K)=\frac{1}{2}{\rm d}^2f/{\rm d}K^2$. In particular, if
$f(K)$ has a maximum at large density or curvature (as it is often exploited
in singularity-free bounce models of loop quantum cosmology), $\beta(K)$ is
negative in this regime.

For instance, in spatially flat isotropic minisuperspace models, $K={\cal H}$
can be taken as the Hubble parameter, and it appears as a connection component
in holonomies (\ref{Hol}) computed along curves in isotropic space. The
boundedness of matrix elements of holonomies is often taken as a motivation to
replace $K^2={\cal H}^2$ in the classical Friedmann equation by a function
$f(K)=\sin^2(\ell K)/\ell^2$, with a length parameter $\ell$ that could be the
Planck length. (With this modification, $K={\cal H}+O(\ell^2)$ with quantum
corrections.)  Although not only the value of $\ell$ but also the precise
functional form of $f(K)$ are subject to quantization ambiguities, which at
present remain poorly controlled, the bounded nature of $f(K)$ is considered
robust because it comes from the boundedness of holonomies for compact
groups. One obtains a modified Friedmann equation \cite{GenericBounce}
\begin{equation} \label{ModFried}
 \frac{\sin^2(\ell K)}{\ell^2}= \frac{8\pi G}{3} \rho
\end{equation}
or, after computing the holonomy corrections by which $K$ differs from ${\cal
  H}$ \cite{AmbigConstr,APSII},
\begin{equation} \label{EffFried}
 {\cal H}^2 = \frac{8\pi G}{3} \rho\left(1-{\textstyle\frac{8}{3}}\pi G\ell^2
   \rho\right)\,. 
\end{equation}
(This equation is an exact effective equation only if the sole matter content
is a free, massless scalar \cite{BouncePert}. In all other cases,
higher-derivative corrections appear which may rival effects from holonomy
modifications \cite{ReviewEff}. So far, no sufficiently general form of higher
time derivatives is known in these models.)

If one takes equation (\ref{ModFried}) at face value, it implies that the
energy density must be bounded as long as $\ell$ is non-zero:
$\rho\leq\rho_{\rm max}=3/(8\pi G\ell^2)$. The version (\ref{EffFried}) of the
equation then suggests that an extremum of the scale factor is reached at the
maximum density $\rho_{\rm max}$ which can be shown to be a minimum. Holonomy
modifications of loop quantum cosmology therefore lead to bounce
models. However, for $f(K)$ of the required form, one obtains (\ref{TTbeta})
with a correction function $\beta(K)=\cos(2\ell K)$ which is negative at high
density around the bounce.

When $\beta$ is different from one, or even negative, the classical relation
(\ref{TT}) is strongly modified. The drastic nature of this quantum correction
can be understood by noting that (\ref{TT}) with the opposite sign is obtained
for deformations of hypersurfaces in Euclidean space rather than Lorentzian
space-time. Bounded curvature as a consequence of holonomy modifications in
loop quantum gravity, as realized in bouncing solutions of (\ref{EffFried}),
can therefore be obtained only at the expense of having signature change at
large curvature \cite{Action,PhysicsToday,SigChange}. No new assumption is
required to arrive at this conclusion; one only combines effective theory with
the general conditions that loop quantum gravity should have some
semiclassical states (in a weak sense since the $\hbar$-order is not fixed)
and that it provides an anomaly-free quantization with a closed algebra of
symmetry generators.

For negative $\beta$, one obtains elliptic mode equations instead of
hyperbolic ones, of the form
\begin{equation} \label{mode}
 -\frac{1}{c^2} \frac{\partial^2u}{\partial t^2}+\beta(K)
 \Delta u=S
\end{equation}
with some source term $S$. In $\beta(K)$, $K$ changes according to a
background solution of (\ref{EffFried}). An initial-value problem is no longer
well-posed; it needs to be replaced by a boundary-value problem in all four
dimensions. If one were to use an initial-value problem even through the
elliptic regime (at high curvature), one would obtain unstable
solutions. Moreover, even a well-posed problem for the mixed-type differential
equations one is dealing with in this context, found by Tricomi in the 1930s,
generically leads to large mode amplitudes somewhere in a region of interest
\cite{Tricomi}. (Mixed-type partial differential equations appear also in
models of transonic flow. Locally large amplitudes in this context are
well-known from sonic booms.) While it is possible to use cosmological
perturbation theory in order to derive the local form of mode equations
(\ref{mode}) including their mixed-type nature \cite{ScalarHolInv}, global
solutions appear to be inconsistent with inhomogeneity being treated as a
perturbation.

These results give rise to two final conclusions. First, as promised, there is
a model in which quantum-field theory on a (bouncing) background gives
drastically different results from a background-independent treatment which
takes into account the symmetries of (quantum) space-time. The former approach
would simply assume that the bouncing background has the standard space-time
structure, and inhomogeneity could evolve through the bounce
perturbatively. Such models have been developed, but even though they are
sometimes thought of as models of quantum gravity (the title of
\cite{AANLetter}, for instance, advertizes ``a quantum gravity extension of
the inflationary scenario'' although it is based on classical assumptions
about inhomogeneity in space-time), they assume classical properties of the
background which are violated in currently known anomaly-free models with the
same quantum corrections in the background dynamics (\ref{EffFried}).

The second conclusion is a sobering note on potential derivations of
observational signatures of loop quantum gravity (while other approaches to
quantum gravity not based on holonomy modifications may have a more positive
outlook). The Planck regime in this theory appears to be too ambiguous to
produce reliable pre-big bang information, and perturbative inhomogeneity is
likely to be inadequate owing to instabilities induced by the very same
quantum-gravity effects that have been exploited for singularity
resolution. One might use the theory for computations of quantum corrections
at sub-Planckian densities, when the curvature parameter $K$ is small enough
for $\beta(K)$ to turn positive. The implication of loop modifications is then
to replace the singular beginning of the expansion phase in the usual
inflationary models by a non-singular beginning marked by the first emergence
of time at the boundary of the Euclidean phase. Additional quantum corrections
characteristic of the loop approach are small in this regime, but they could
potentially be computed by effective methods and, once they are known in
detail and with a classification of all possible quantum ambiguities within
this theory, compared with observations. A key new effect, directly related to
non-classical space-time structures, is related to modified propagation speeds
of various modes implied by (\ref{mode}) for $0<\beta\not=1$. Importantly,
scalar and tensor modes may have different speeds \cite{InflConsist}, so that
characteristic modifications of the usual scalar-to-tensor ratio can
result. The theory might have something to say about preferred initial states
of an inflaton (see for instance \cite{HolState}), but this issue is still
being analyzed.

To conclude, surprising and perhaps useful effects have been found in models
of loop quantum gravity. In scenarios of structure formation, the consequences
are not always as they first appear in the simplest (homogeneous) models which
by necessity are blind to the underlying quantum space-time structure. There
remain several difficult questions to be addressed, for instance about
anomalies, before one can make reliable predictions. Nevertheless, at a
conceptual level the question of how loop quantum gravity could be combined
with the inflationary scenario is producing several interesting lessons.

\bigskip

Some of the research reported here was supported by NSF grant PHY-1307408. The
author is grateful to Jean-Philippe Uzan for an invitation to contribute this
review to a special issue of {\em Comptes Rendus Physique}, published by the
French Academy of Sciences.



\begin{thebibliography}{10}

\bibitem{EstQG}
L.~M.\ Krauss and F.\ Wilczek,
\newblock Using cosmology to establish the quantization of gravity,
\newblock {\em Phys.\ Rev.\ D} 89 (2014) 046501, [arXiv:1309.5343]

\bibitem{GCPheno}
G.\ Calcagni,
\newblock Observational Effects from Quantum Cosmology,
\newblock {\em Annalen Phys.} 525 (2012) 323--338; A165, [arXiv:1209.0473]

\bibitem{ObsLQC}
A.\ Barrau, T.\ Cailleteau, J.\ Grain, and J.\ Mielczarek,
\newblock Observational issues in loop quantum cosmology,
\newblock {\em Class.\ Quant.\ Grav.} 31 (2014) 053001, [arXiv:1309.6896]

\bibitem{EffectiveGR}
J.~F.\ Donoghue,
\newblock General relativity as an effective field theory: The leading quantum
  corrections,
\newblock {\em Phys.\ Rev.\ D} 50 (1994) 3874--3888, [gr-qc/9405057]

\bibitem{BurgessLivRev}
C.~P.\ Burgess,
\newblock Quantum Gravity in Everyday Life: General Relativity as an Effective
  Field Theory,
\newblock {\em Living Rev.\ Relativity} 7 (2004), [gr-qc/0311082],
\newblock http://www.livingreviews.org/lrr-2004-5

\bibitem{ADM}
R.\ Arnowitt, S.\ Deser, and C.~W.\ Misner,
\newblock The Dynamics of General Relativity, In L.\ Witten, editor, {\em
  Gravitation: An Introduction to Current Research},
\newblock Wiley, New York, 1962,
\newblock Reprinted in \cite{ADMRe}

\bibitem{CUP}
M.\ Bojowald,
\newblock {\em Canonical Gravity and Applications: Cosmology, Black Holes, and
  Quantum Gravity},
\newblock Cambridge University Press, Cambridge, 2010

\bibitem{Regained}
S.~A.\ Hojman, K.\ Kucha\v{r}, and C.\ Teitelboim,
\newblock Geometrodynamics Regained,
\newblock {\em Ann.\ Phys.\ (New York)} 96 (1976) 88--135

\bibitem{ConsAlgebroid}
C.\ Blohmann, M.~C.\ Barbosa~Fernandes, and A.\ Weinstein,
\newblock Groupoid symmetry and constraints in general relativity. 1:
  kinematics,
\newblock {\em Commun.\ Contemp.\ Math.} 15 (2013) 1250061, [arXiv:1003.2857]

\bibitem{EffConsQBR}
M.\ Bojowald and S.\ Brahma,
\newblock Effective constraint algebras with structure functions,
  [arXiv:1407.4444]

\bibitem{LoopRep}
C.\ Rovelli and L.\ Smolin,
\newblock Loop Space Representation of Quantum General Relativity,
\newblock {\em Nucl.\ Phys.\ B} 331 (1990) 80--152

\bibitem{ALMMT}
A.\ Ashtekar, J.\ Lewandowski, D.\ Marolf, J.\ Mour\~ao, and T.\ Thiemann,
\newblock Quantization of Diffeomorphism Invariant Theories of Connections with
  Local Degrees of Freedom,
\newblock {\em J.\ Math.\ Phys.} 36 (1995) 6456--6493, [gr-qc/9504018]

\bibitem{AnoFree}
T.\ Thiemann,
\newblock Anomaly-Free Formulation of Non-Perturbative,
  Four-Dimensional Lorentzian Quantum Gravity,
\newblock {\em Phys.\ Lett.\ B} 380 (1996) 257--264, [gr-qc/9606088]

\bibitem{EffAc}
M.\ Bojowald and A.\ Skirzewski,
\newblock Effective Equations of Motion for Quantum Systems,
\newblock {\em Rev.\ Math.\ Phys.} 18 (2006) 713--745, [math-ph/0511043]

\bibitem{EffCons}
M.\ Bojowald, B.\ Sandh\"ofer, A.\ Skirzewski, and A.\ Tsobanjan,
\newblock Effective constraints for quantum systems,
\newblock {\em Rev.\ Math.\ Phys.} 21 (2009) 111--154, [arXiv:0804.3365]

\bibitem{JR}
J.~D.\ Reyes,
\newblock {\em Spherically Symmetric Loop Quantum Gravity: Connections to
  2-Dimensional Models and Applications to Gravitational Collapse},
\newblock PhD thesis, The Pennsylvania State University, 2009

\bibitem{ScalarHolInv}
T.\ Cailleteau, L.\ Linsefors, and A.\ Barrau,
\newblock Anomaly-free perturbations with inverse-volume and holonomy
  corrections in Loop Quantum Cosmology,
\newblock {\em Class.\ Quantum Grav.} 31 (2014) 125011, [arXiv:1307.5238]

\bibitem{HigherSpatial}
M.\ Bojowald, G.~M.\ Paily, and J.~D.\ Reyes,
\newblock Discreteness corrections and higher spatial derivatives in effective
  canonical quantum gravity,
\newblock {\em Phys.\ Rev.\ D} 90 (2014) 025025, [arXiv:1402.5130]

\bibitem{DeformedCosmo}
A.\ Barrau, M.\ Bojowald, G.\ Calcagni, J.\ Grain, and M.\ Kagan,
\newblock Anomaly-free cosmological perturbations in effective canonical
  quantum gravity,
\newblock {\em JCAP} 05 (2015) 051, [arXiv:1404.1018]

\bibitem{GenericBounce}
G.\ Date and G.~M.\ Hossain,
\newblock Genericity of Big Bounce in isotropic loop quantum cosmology,
\newblock {\em Phys.\ Rev.\ Lett.} 94 (2005) 011302, [gr-qc/0407074]

\bibitem{AmbigConstr}
K.\ Vandersloot,
\newblock On the Hamiltonian Constraint of Loop Quantum Cosmology,
\newblock {\em Phys.\ Rev.\ D} 71 (2005) 103506, [gr-qc/0502082]

\bibitem{APSII}
A.\ Ashtekar, T.\ Pawlowski, and P.\ Singh,
\newblock Quantum Nature of the Big Bang: Improved dynamics,
\newblock {\em Phys.\ Rev.\ D} 74 (2006) 084003, [gr-qc/0607039]

\bibitem{BouncePert}
M.\ Bojowald,
\newblock Large scale effective theory for cosmological bounces,
\newblock {\em Phys.\ Rev.\ D} 75 (2007) 081301(R), [gr-qc/0608100]

\bibitem{ReviewEff}
M.\ Bojowald,
\newblock Quantum Cosmology: Effective Theory,
\newblock {\em Class.\ Quantum Grav.} 29 (2012) 213001, [arXiv:1209.3403]

\bibitem{Action}
M.\ Bojowald and G.~M.\ Paily,
\newblock Deformed General Relativity and Effective Actions from Loop Quantum
  Gravity,
\newblock {\em Phys.\ Rev.\ D} 86 (2012) 104018, [arXiv:1112.1899]

\bibitem{PhysicsToday}
M.\ Bojowald,
\newblock Back to the beginning of quantum spacetime,
\newblock {\em Physics Today} 66 (2013) 35

\bibitem{SigChange}
J.\ Mielczarek,
\newblock Signature change in loop quantum cosmology,
\newblock {\em Springer Proc.\ Phys.} 157 (2014) 555, [arXiv:1207.4657]

\bibitem{Tricomi}
F.~G.\ Tricomi,
\newblock {\em Repertorium der Theorie der Differentialgleichungen},
\newblock Springer Verlag, 1968

\bibitem{AANLetter}
I.\ Agull\'o, A.\ Ashtekar, and W.\ Nelson,
\newblock A Quantum Gravity Extension of the Inflationary Scenario,
\newblock {\em Phys.\ Rev.\ Lett.} 109 (2012) 251301, [arXiv:1209.1609]

\bibitem{InflConsist}
M.\ Bojowald, G.\ Calcagni, and S.\ Tsujikawa,
\newblock Observational constraints on loop quantum cosmology,
\newblock {\em Phys.\ Rev.\ Lett.} 107 (2011) 211302, [arXiv:1101.5391]

\bibitem{HolState}
J.\ Mielczarek,
\newblock Inflationary power spectra with quantum holonomy corrections,
\newblock {\em JCAP} 03 (2014) 048, [arXiv:1311.1344]

\bibitem{ADMRe}
R.\ Arnowitt, S.\ Deser, and C.~W.\ Misner,
\newblock The Dynamics of General Relativity,
\newblock {\em Gen.\ Rel.\ Grav.} 40 (2008) 1997--2027

\end{thebibliography}

\end{document}